\documentclass[aps,prb,showpacs,twocolumn,superscriptaddress]{revtex4-2}
\usepackage[english]{babel}
\usepackage[T2A]{fontenc}
\usepackage[koi8-r]{inputenc}
\usepackage{amssymb,amsmath,amsfonts,amsthm}
\usepackage{mathrsfs}
\usepackage[mathcal]{eucal}

\usepackage[dvips]{graphicx}
\usepackage[colorlinks]{hyperref}

\usepackage{chngcntr}

\begin{document}

\title{Effect of thermal fluctuations on topological crossover in the chiral 
$d+id$ superconducting phase}

\author
{A.G.~Groshev}
\email{groshev_a.g@mail.ru}
\affiliation{Udmurt Federal Research Center of the Ural Branch of the 
Russian Academy of Sciences, T. Baramzinoy st. 34, Izhevsk 426067, Russia}
\author
{A.K.~Arzhnikov}
\email{arzhnikof@bk.ru}
\address{Udmurt Federal Research Center of the Ural Branch of the Russian 
Academy of Sciences, T. Baramzinoy st. 34, Izhevsk 426067, Russia}

\date{\today}

\begin{abstract}

The effect of thermal fluctuations on the temperature dependence of the topological 
index $C_1$ of the chiral $d+id$ superconducting  phase of a two-dimensional single-band 
model on a triangular lattice is investigated. Thermal fluctuations are taken into 
account within the framework of the self-consistent functional-integral theory. 
It is established that when the nodal points are located far inside (outside) the Fermi 
contour of the normal phase,  thermal fluctuations expand the relative temperature ranges 
in which the values of the topological index are close to integer values $C_1$= 4(-2). This  
expansion depends both on the value of the topological index and on the magnitude of the 
effective attraction between the electrons.  However, as the nodal points approach the Fermi 
contour, topological crossovers to new $C_1$ values are observed, which can persist over a 
wide temperature range. The nature and degree of influence of thermal fluctuations on these 
crossovers are established. It is assumed that the observed effects may also manifest in the 
edge state behavior of a similar system with open boundaries.
\end{abstract}


\maketitle


The study of superconducting states with nontrivial topology in condensed matter physics 
is motivated by fundamental questions regarding the emergence of Majorana states in systems. 
These states have non-Abelian statistics and have potential applications in fault-tolerant 
quantum computers, in the implementation of high-speed information transmission, and spintronics 
(see, for example, \cite{Nayak_2008,Zeng_2019,Valkov_2022,Qi_2011}). One of the research directions 
in this area is related to the topological properties of superconducting systems taking into account 
real conditions, i.e. when energy fluctuations exist, there are no conservation laws for the number 
of particles, etc. This forces us to consider effective non-Hermitian Hamiltonians, which 
assume the decay of quantum states. Non-Hermitian Hamiltonians are also used to describe 
some forms of dissipation in quantum many-particle and disordered systems. Naturally, the 
study of the topological properties of superconducting systems at finite temperatures becomes 
relevant \cite{Groshev_2024,Hastings_2011,Unanyan_2020}. When the temperature changes, the 
system can maintain topological properties over a wide temperature range or move from one 
non-trivial topological phase to another. These issues have important fundamental and practical 
significance (see, for example, 
\cite{Yuto_2020,Kawabata_2019,Long_2022,Markov_2021,Li_2023,Niu_1985,Valkov_2019}).

The topological properties of superconducting phases in two-dimensional materials at zero 
temperature are characterized by integer values of topological index (TI), which is 
expressed in terms of single-particle Green's function \cite{Valkov_2019}. A change in these 
values signals topological phase transitions. A similar TI is used to characterize two-dimensional 
topological insulators and the integer quantum Hall effect \cite{Ishikawa_1987}. When considering 
Green's functions in models without interaction or in the approximation of a mean field with a 
frequency-independent self-energy part \cite{Rachel_2010,Raghu_2008}, the expression for TI can 
be explicitly frequency-integrated. In this case, TI reduces to the well-known definition of the 
winding number associated with the Chern number and Berry phases in momentum space 
\cite{Valkov_2019,Niu_1985}, and topological phase transitions are determined by a change in the 
number of zeros of the complex gap function in the Brillouin zone (nodal points) located inside 
the Fermi contour of the normal phase. In the general case, in addition to nodal points, TI is 
determined by the zeros (poles) of the one-electron Green's function, i.e., the poles in the energy 
dependence of the self-energy part, which can occur when many-body effects, thermal fluctuations, 
and disorder are taken into account \cite{Budich_2012,Wang_2011,Nagai_2020,Zheng_2019}. In this case, 
the system under consideration is described by non-Hermitian effective Hamiltonian with a complex 
self-energy part. In this paper, the temperature behavior of TI is considered when thermal 
fluctuations of the superconducting order parameter are taken into account.

The effect of thermal fluctuations on a temperature behavior of the topological index 
$C_1$ of the chiral $d+id$ superconducting phase of the two-dimensional single-band model on a 
triangular lattice, taking into account the nearest neighbors in the pairing channel, has already 
been studied in \cite{Groshev_2024}. However, the TI analysis showed that the emerging pole structure 
in the energy dependence of the self-energy part does not lead to new topological transitions and 
the temperature behavior of TI is completely determined by the root-mean-square of phase fluctuations 
(RMSF) of the superconducting order parameter. It should also be noted that the model used in 
\cite{Groshev_2024} has only one nodal point (in the center of the Brillouin zone), which can 
be located on the Fermi contour only at the limiting value of the electron concentration $n=0$. 
However, in this range, the ground state of the system has $s$-symmetry \cite{Groshev_2024} and 
is topologically trivial. In the range of electronic concentrations where the $d+id$ phase is 
implemented, this nodal point corresponds to a single topological phase of the model with a TI value of 
$C_1=-2$.

At the same time, for the formation of edge states, including for the Majorana mode, it turns 
out to be important on whether nodal points (with quasi-particle dispersion of the Dirac type) are 
located near the Fermi surface (see, for example, \cite{He_2025}). From this point of view, an 
important result was obtained in \cite{Groshev_2025}, in which, within the framework of the 
Hartree-Fock (HF) approximation, the temperature behavior of the topological index $C_1$ of the 
chiral $d+id$ superconducting phase of a two-dimensional single-band model on a triangular lattice 
was investigated, taking into account the second nearest neighbors in the pairing channel. In this 
case, there may be several nodal points inside the Fermi contour that intersect the Fermi contour 
when the concentration of charge carriers changes. The phase diagram (dependence of the topological 
index $C_1$ on temperature and chemical potential) of this model in the HF approximation is shown 
in Fig.~1. 
\begin{figure}[t!]
\begin{center}
\includegraphics[scale=0.52]{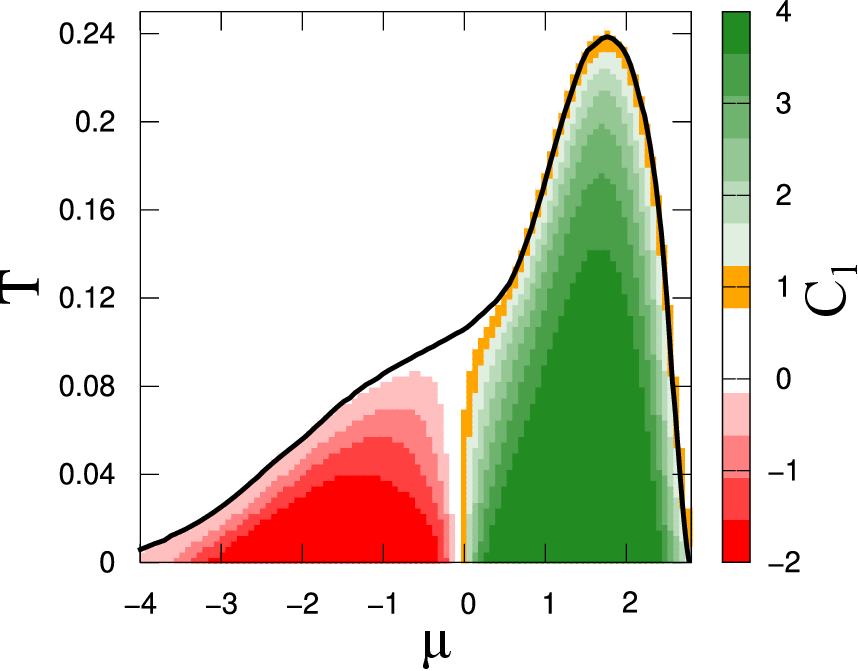}
\caption{(Color online) The dependence of the topological index $C_1$ on temperature and chemical 
potential (phase diagram) in the HF approximation at the value of the interelectron attraction 
parameter $V=2$. The temperature of the superconducting transition $T_c$ is represented by a black 
line.}
\end{center}
\end{figure}
When the nodal points are located far inside (outside) the Fermi contour, the values of the topological 
index over a wide temperature range are close to $C_1= 4(-2)$. However, when the nodal points approached 
the Fermi contour these values were preserved only at low temperatures,  while values close to $C_1=1$ 
were realized over a wide temperature range. Thus, a topological crossover was observed from the values 
of TI $C_1=4$ and $C_1=-2$ to a value close to $C_1=1$. We note that the value of TI $C_1=1$ in the ground 
state was obtained in \cite{Zhou_2008} via direct integration along contours passing enclosing six nodal 
points located on the Fermi contour. The temperature dependence of TI obtained in 
\cite{Groshev_2025} is explained by the temperature dependence of the probabilities of filling nodal 
points. The study in \cite{Groshev_2025} did not take into account thermal fluctuations,  as it was assumed 
that taking them into account would not change the qualitative behavior of the detected topological 
crossover, at least for a sufficiently small amount effective electron attraction, as long 
as the renormalization of the Fermi contour by thermal fluctuations remains insignificant. In this paper, 
in contrast to \cite{Groshev_2025}, the behavior of the topological crossover is investigated when thermal 
fluctuations are taken into account for both small and sufficiently large electron attraction, i.e. in 
situations with  weak and strong renormalization of the Fermi contour by thermal fluctuations.

\section{Model and results}

A two-dimensional model of a system with an effective attraction of electrons located at the 
sites of a triangular lattice with a Hamiltonian 
\begin{equation}
\label{eq:hamiltonian1}
{\hat {\cal H}}=\sum_{i,j,s}t_{ij}{\hat c}_{is}^{+}{\hat c}_{js}^{}-
\sum_{j}\mu{\hat n}_{j}-
V\sum_{j,\delta}{\hat n}_{j\uparrow}{\hat n}_{j+\delta\downarrow},
\end{equation} 
is considered, where $t_{ij}=-t$-matrix elements of electron hopping to the nearest sites; 
$\hat{c}_{js}^{+}(\hat{c}_{js})$-electron creation (unnihilation) operators at site $j$ with 
spin projection $s$; $n_{js}=\hat{c}_{js}^{+}\cdot\hat{c}_{js}$-electron number operator at 
site $j$ with spin projection $s$; $\delta$-the second nearest neighbors of site $j$; $n_{j}$ 
is the operator of the number of electrons at the site $j$; $\mu$ is the chemical potential; 
$V$ is the parameter of the interelectron attraction within the second coordination sphere. 
The basic spin-singlet superconducting state with a nontrivial chiral $d+id$ symmetry of the 
order parameter of such a system has been extensively studied \cite{Zhou_2008,Valkov_2019,Kumar_2003}. 
Similar Hamiltonians have been used to describe the properties of real superconductors, such 
as water-intercalated layered sodium cobaltites Na$_{x}$CoO$_{2}\cdot y$H$_{2}$O \cite{Zhou_2008}. 
According to the authors \cite{Zhou_2008},  the inclusion of the second nearest neighbors in 
the pairing channel is dominant in these compounds, and the presence of a topologically nontrivial 
point at the Fermi level makes it possible to explain the temperature dependences of the spin 
relaxation rate and the Knight shift in NMR experiments \cite{Fujimoto_2004,Zheng_2006}.

In the case of superconducting phases with a nontrivial topology, it is generally accepted to 
calculate TI in terms of a single-particle Green's function with imaginary time 
\cite{Volovik_1989,Ishikawa_1987}. The expression for TI in terms of retarded and advanced 
single-particle Green's functions, valid at finite temperatures, was derived in \cite{Groshev_2024} 
by taking into account thermal fluctuations in the framework of the self-consistent functional-integral 
theory directly from the Kubo formula for conductivity in the following form: 
\begin{equation}
\label{eq:Sigma_H_T}
\begin{array}{c}
\displaystyle
C_{1}=\int\limits_{-\infty}^{+\infty}dE
\int\limits_{-\pi}^{\pi}\int\limits_{-\pi}^{\pi} \frac{dk_{1}dk_{2}}{16\pi^{2}}
th\left(\frac{\beta E}{2}\right)
\times
\\
\displaystyle
\times
tr\left[\partial_{k_{1}}G^{-1}_{}(E)K^{-}(E)\partial_{k_{2}}G^{-1}_{}(E)\partial_{E}K^{+}(E)
-
\right.
\\
\displaystyle
\left.
-K^{-}(E)\partial_{k_{1}}G^{-1}_{}(E)\partial_{E}K^{+}(E)\partial_{k_{2}}G^{-1}_{}(E)\right].
\end{array}
\end{equation} 
Here the notation $K^{\pm}(E)=$ $[F^{A}(E)\pm F^{R}(E)]/2$ is used, where the retarded (advanced) 
$F^{R(A)}(E)$ the one-particle Green's operator is the resolvent of the Hamiltonian 
(\ref{eq:hamiltonian1}) averaged over thermal fluctuations $\Delta U$ in the HF approximation 
${\cal H}={\cal H}_{HF}+\Delta U$, $F^{R(A)}(E)=$
$\lim_{z\to E\pm i0}\left\langle[z-{\cal H}_{HF}-\Delta U]^{-1}\right\rangle=
[E-{\cal H}_{HF}-\Sigma^{R(A)}(E)]^{-1}$, $G(E)=[E-{\cal H}_{HF}]^{-1}$ is a one-particle Green's 
operator in the HF approximation, $\partial_{E}=\partial/\partial_{E}$, $\beta =1/k_{B}T$. 
Following the approach of \cite{Groshev_2024}, when calculating the self-energy part, we restrict 
ourselves quadratic approximation in the fluctuating potential $\Delta U$. Without taking into 
account thermal fluctuations, the self-energy part $\Sigma^{R(A)}(E)$ of the Green's functions 
$F^{A(R)}(E)$ vanishes. In this case, expression (\ref{eq:Sigma_H_T}) reduces to the expression for 
the winding number associated with the Berry phase and the Chern number \cite{Valkov_2019}. 

Calculations of the temperature dependence of TI without thermal fluctuations were carried out in 
\cite{Groshev_2025}. In this paper, when calculating TI, we take into account thermal fluctuations, 
therefore, expression (\ref{eq:Sigma_H_T}) is used in all calculations of the temperature dependence 
of TI. Since we do not discuss the behavior of real systems, all energy quantities are presented 
in units of the hopping integral $t$. Taking into account thermal fluctuations leads to the appearance 
of the self-energy part in the Green's functions and, consequently, to the renormalization of the Fermi 
contour. In our case, it is convenient to write this renormalization in the form of a renormalized 
chemical potential $\mu^{\ast }$, which is determined by the expression 
$\mu^{\ast }=\mu-0.5\left(\Sigma^{\uparrow}(0)-\Sigma^{\downarrow}(0)\right)$, where $\mu$ is the 
chemical potential without taking into account thermal fluctuations; $\Sigma^{\uparrow(\downarrow)}(0)$ 
are the diagonal components of the operator of the self-energy part. If the nodal points are located 
sufficiently far from the Fermi contour and do not intersect the renormalized Fermi contour as the 
temperature changes, then thermal fluctuations have little effect on the TI across its entire range 
of stability, except at high temperatures (see \cite{Groshev_2024}). 
At the same time, the six nodal points inside the Fermi contour are located much closer to it for the 
topological phase with a value of $C_{1}=4$ than the single nodal point in the center of the Brillouin 
zone for the topological phase $C_{1}=-2$. Consequently, the  $C_{1}=4$ phase is more sensitive to 
thermal fluctuations. When the nodal points are located near the Fermi contour, their intersection with 
the renormalized Fermi contour becomes possible with a change in temperature. In this case, there are 
two regimes of renormalization of the chemical potential: a strong and a weak renormalization regime. 
To demonstrate these two regimes, we performed TI calculations for the values $V=2$ and $V=4$ (the value 
of the parameter $V=2$ corresponds to the choice made in \cite{Groshev_2025}).

\begin{figure}[t!]
\begin{center}
\includegraphics[scale=0.43]{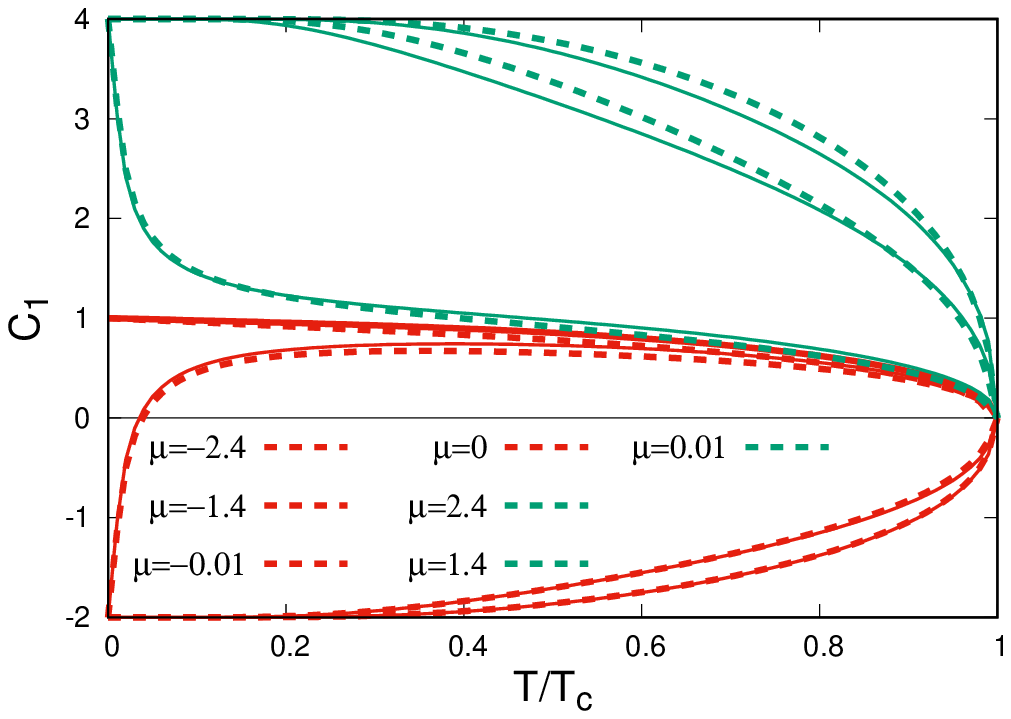}
\caption{(Color online) Temperature dependences of the topological index $C_{1}$ in 
the HF approximation for different values of the interelectron attraction parameter 
$V=2$ (dotted lines) and $V=4$ (solid lines).}
\end{center}
\end{figure}
The calculated temperature dependences of TI without taking into account thermal fluctuations 
at the values of the interelectron attraction parameter $V=2$ and $V=4$ are shown in Fig.~2. It 
can be seen from the figure that, despite the significant difference in the temperature values of 
the superconducting transition $T_c$ for $V=2$ and $V=4$, the corresponding values of TI in the 
relative variables $T/T_c$ differ slightly from each other. 
\begin{figure}[t!]
\begin{center}
\includegraphics[scale=0.43]{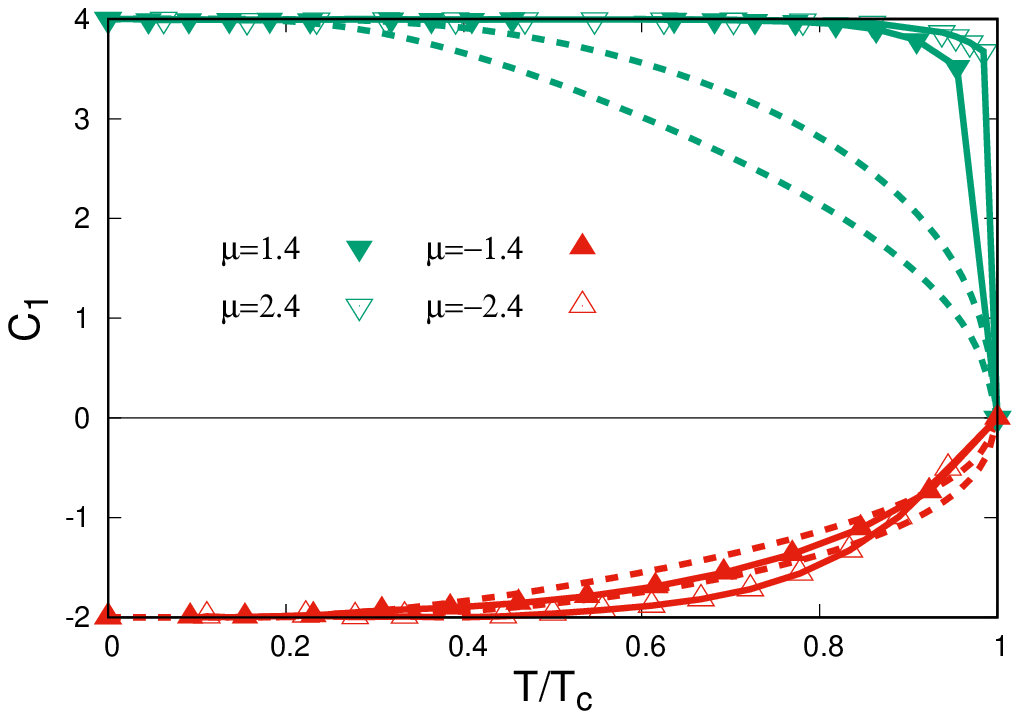}
\caption{(Color online) Temperature dependences of the topological index $C_{1}$ in the 
HF approximation (dotted lines) and taking into account thermal fluctuations (solid lines) 
at the value of the interelectron attraction parameter $V=2$.}
\end{center}
\end{figure}
\begin{figure}[t!]
\begin{center}
\includegraphics[scale=0.43]{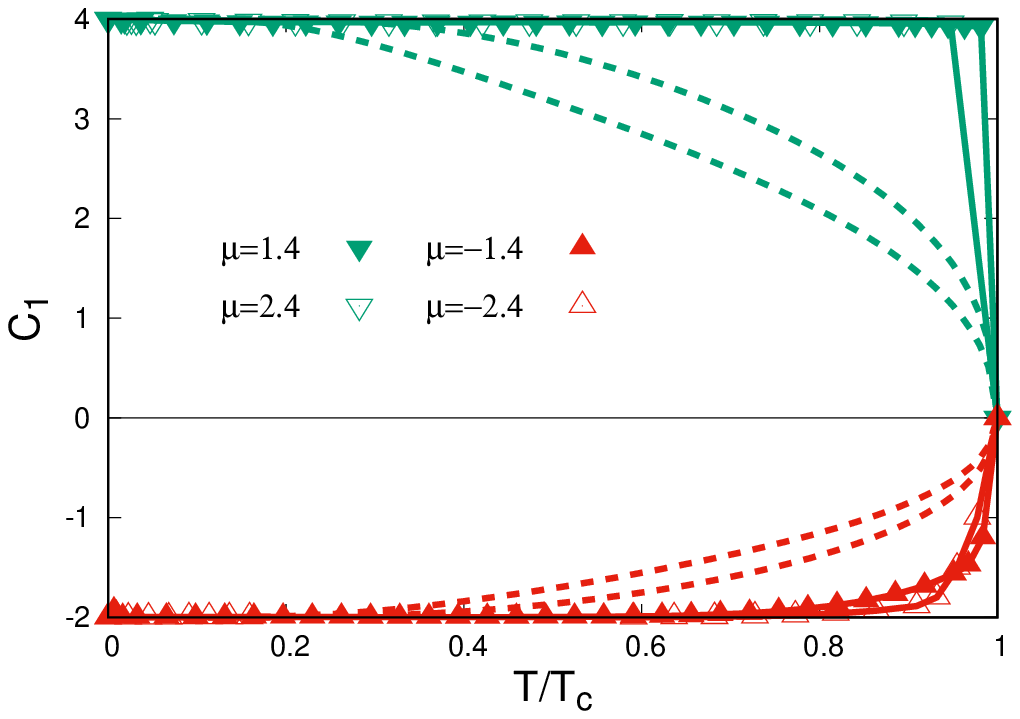}
\caption{(Color online) Temperature dependences of the topological index $C_{1}$ in the 
HF approximation (dotted lines) and taking into account thermal fluctuations (solid lines) 
at the value of the interelectron attraction parameter $V=4$.}
\end{center}
\end{figure}
A comparison of the temperature dependences of TI with and without thermal fluctuations in conditions 
when the nodal points are far from the Fermi contour of the normal phase is shown in Fig.~3 for $V=2$ 
and in Fig.~4 for $V=4$. These figures demonstrate that accounting for thermal fluctuations leads to 
an expansion of the relative temperature ranges in which the values of the topological index are close 
to the integer values $C_1=4$ and $C_1=-2$. This effect is more pronounced for the $C_1=4$ phase, 
where the nodal  points are located inside the Fermi contour. We have previously results earlier in 
\cite{Groshev_2024}, which considered interelectron attraction was taken into account within the first 
coordination sphere. The expansion of the temperature ranges of TI stability is fully explained by the 
temperature behavior of the RMSF values of the order parameter, which describes the decoherence of the 
propagation of Cooper pairs. The behavior of RMSF is shown in Fig.~5 and Fig.~6 for $V=2$ and $V=4$, 
respectively. The RMSF curves for $V=2$ (Fig.~5)  exhibit a more rapid temperature dependence than 
those for $V=4$  (Fig.~6). Consequently, for a given  value of $V=2$, the decoherence level required 
to suppress topological order is reached at a lower relative temperature ($T/T_c$) compared to the 
case of $V=4$. This results in a broader relative temperature range over which the TI remains stable 
for $V=4$. This broader stability range for $V=4$ is particularly evident for the chemical potential 
values $\mu=-1.4$ and $\mu=-2.4$, which correspond to cases where the nodal points lie outside the 
Fermi contour of the normal phase. These curves are shown in Fig.~3 and Fig.~4 with and without thermal 
fluctuations.
\begin{figure}[t!]
\begin{center}
\includegraphics[scale=0.43]{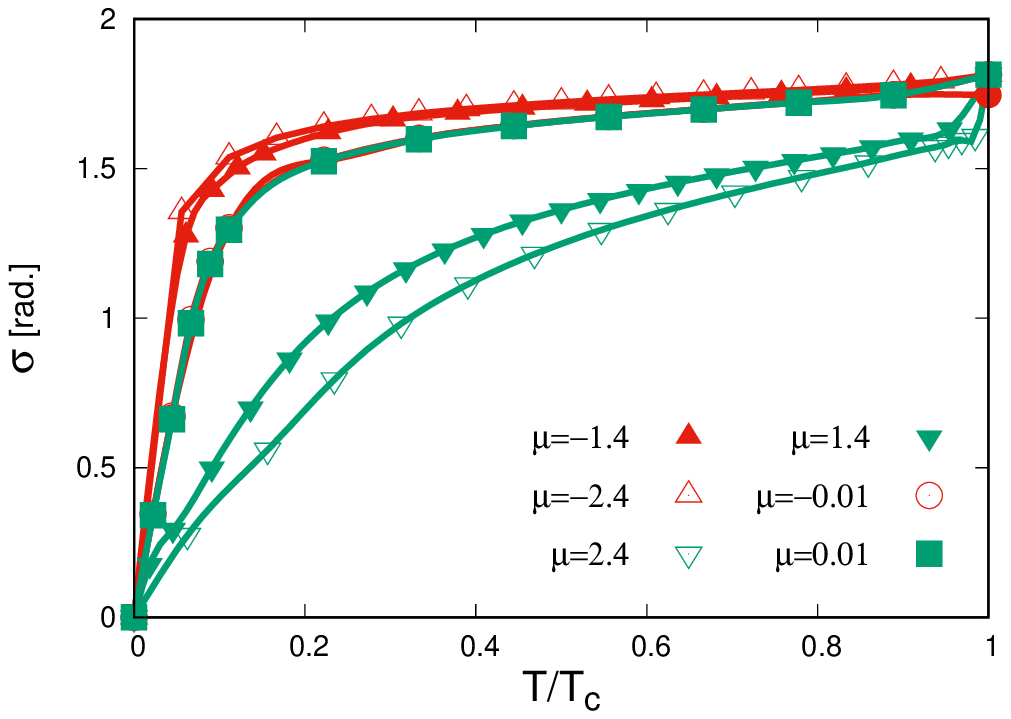}
\caption{(Color online) The temperature behavior of the RMSF value of the superconducting order 
parameter at the value of the interelectron attraction parameter $V=2$.}
\end{center}
\end{figure}
\begin{figure}[t!]
\begin{center}
\includegraphics[scale=0.43]{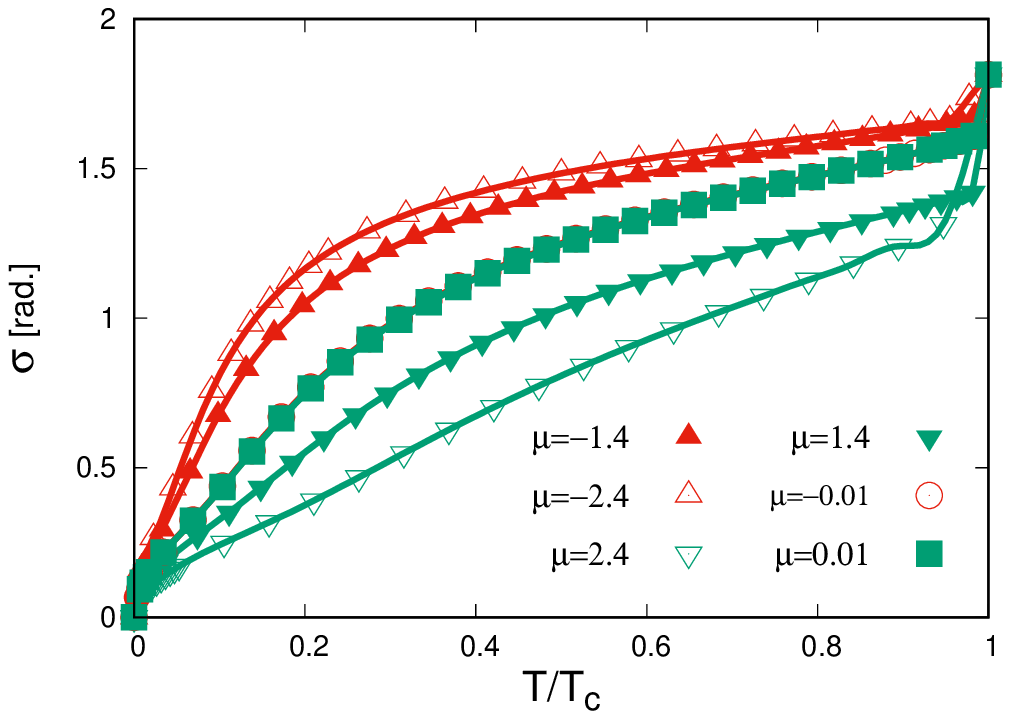}
\caption{(Color online) The temperature behavior of the RMSF value of the superconducting order 
parameter at the value of the interelectron attraction parameter $V=4$.}
\end{center}
\end{figure}

The most interesting and important comparison involves the temperature dependences of TI with and 
without thermal fluctuations under conditions where the nodal points are close to the Fermi 
contour of the normal phase. These dependences are shown in Fig.~7 for $V=2$ and in Fig.~8 for $V=4$.
\begin{figure}[t!]
\begin{center}
\includegraphics[scale=0.43]{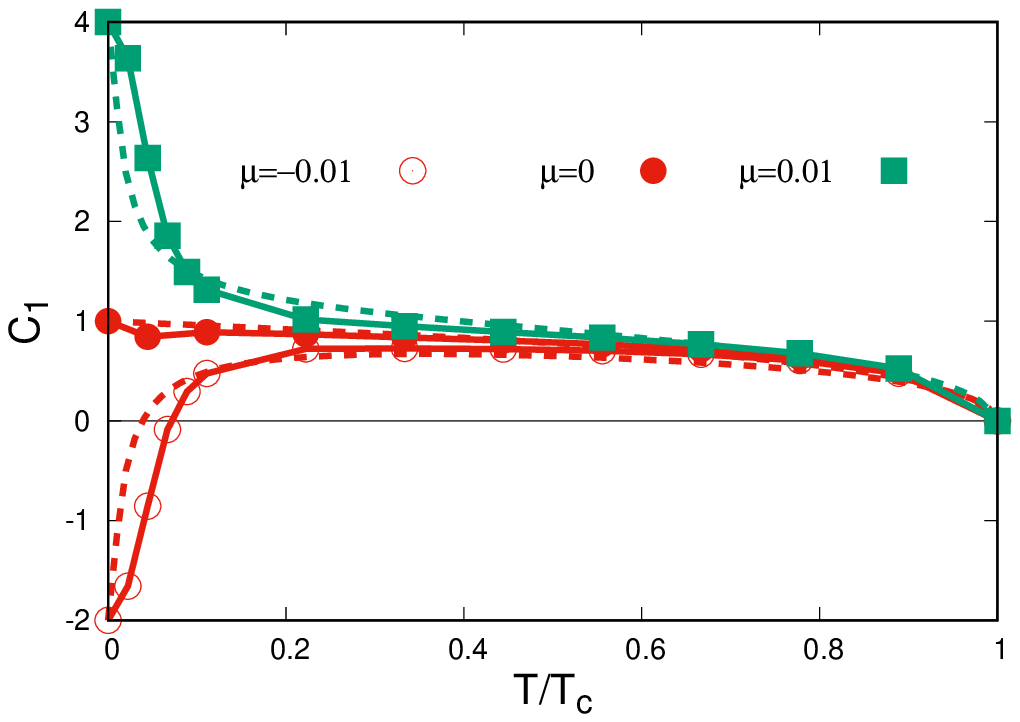}
\caption{(Color online) Temperature dependences of the topological index $C_1$ in the HF 
approximation (dotted lines) and taking into account thermal fluctuations (solid lines) at 
the value of the interelectron attraction parameter $V=2$.}
\end{center}
\end{figure}
\begin{figure}[t!]
\begin{center}
\includegraphics[scale=0.43]{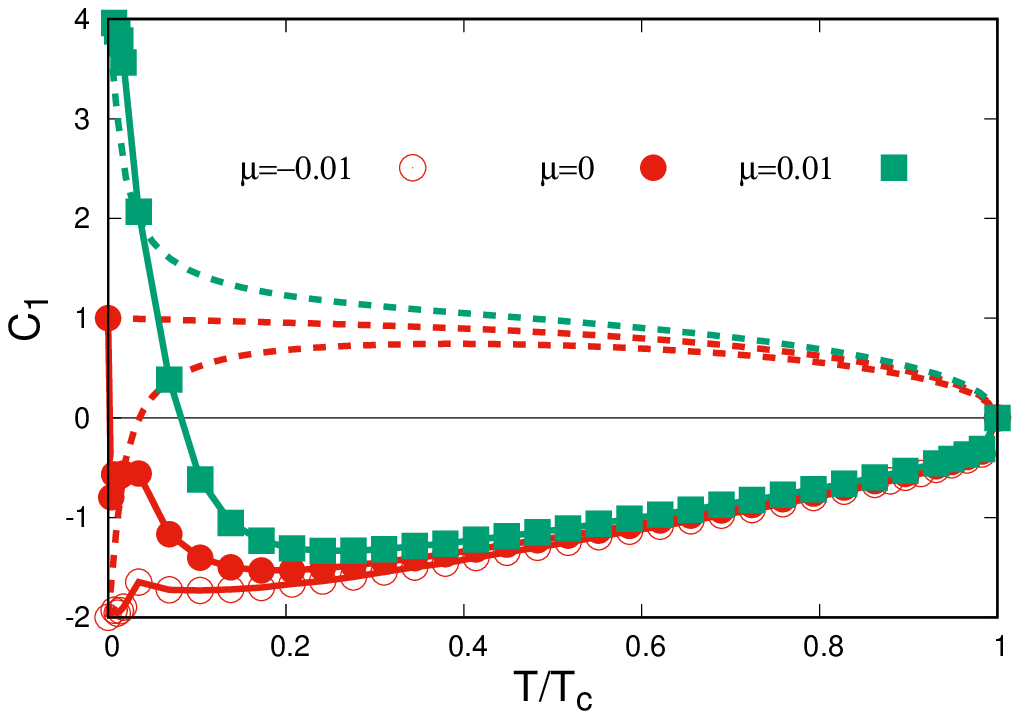}
\caption{(Color online) Temperature dependences of the topological index $C_1$ in the HF 
approximation (dotted lines) and taking into account thermal fluctuations (solid lines) at 
the value of the interelectron attraction parameter $V=4$.}
\end{center}
\end{figure}
In this regime, both strong and weak renormalization of the Fermi contour by thermal fluctuations are 
possible, as illustrated by the temperature dependence of $\mu^{\ast}$ in Fig.~9. For $V=2$, all curves 
in Fig.~9 have weak dependence on temperature up to the temperature of the superconducting transition. 
Nevertheless, at $\mu=0.01$, the nodal points intersect the renormalized Fermi contour and are outside 
it in the high-temperature region at $T/T_c\simeq 0.6$. However, in this temperature region, a topological 
crossover of TI to values $C_1=1$ is already being realized. Consequently, the temperature dependences of 
TI, taking into account thermal fluctuations in Fig.~7, differ slightly from those calculated in the HF 
approximation. In this case, we can talk about the weak influence of thermal fluctuations on the topological 
crossover. A drastic change occurs for $V=4$, where the renormalization leads to a strong temperature 
dependence of the Fermi  contour for all $\mu^{\ast}$ curves  in Fig.~9.  For $V=4$, the nodal points 
intersect the Fermi contour at $\mu =0.01$ already in the low-temperature region $T/T_c\simeq 0.04$. 
Subsequently, in the high-temperature region, the nodal  points lie far outside the Fermi  contour for 
all $\mu$, a region where the TI typically assumes negative values (as seen in Figs.~2-4). As a 
result of this evolution, a topological crossover is observed to a new value of $C_1=-2$, and not to the 
value of $C_1=1$. This topological crossover can occur more sharply at high values of $V$, when the renormalized 
chemical potential $\mu^{\ast}$ has a sharper temperature dependence, and the intersection of nodal points 
by the Fermi contour occurs at lower temperatures. In this case, values close to $C_1=-2$ can be maintained 
over a wide temperature range, and the change in TI resembles the temperature behavior of TI in a situation 
where nodal points are located far outside the Fermi contour. The proximity of the $\mu^{\ast}$ curves in 
the high-temperature region in Fig.~8 are explained by a small difference in the values of $\mu $ 
($\Delta\mu\simeq 0.01$). Note that the intersection of nodal points by the Fermi contour with a temperature 
change can also occur as a result of thermal broadening of the Fermi distribution function, i.e. without 
taking into account thermal fluctuations. As shown in Tables~1 and 2, which list  the characteristic values 
of the chemical potential $\mu $ corresponding to the concentrations of charge carriers $n$ at temperatures 
$T=0$ and $T=T_c$ for realization this mechanism requires values of $\mu\simeq 0.001$ that are closer to 
the Fermi contour. Furthermore, this scenario requires the charge carrier concentration to be held constant 
as the temperature varies experimentally. The weak temperature dependence of the chemical potential $\mu$ 
under this  condition (see Tables~1 and 2) would then lead to a situation analogous to that shown in Fig.~7, 
thereby preserving the topological crossover.

\begin{figure}[t!]
\begin{center}
\includegraphics[scale=0.43]{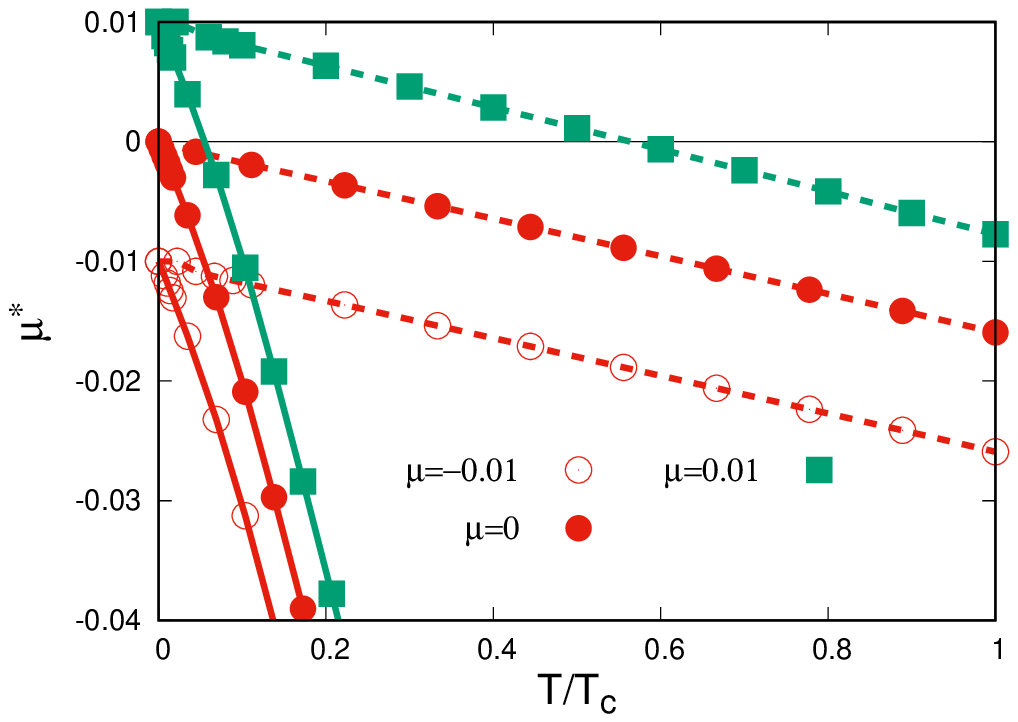}
\caption{(Color online) Temperature dependence of the renormalized chemical potential 
$\mu^{\ast }=\mu-0.5\left(\Sigma^{\uparrow}(0)-\Sigma^{\downarrow}(0)\right)$ at different 
values of the interelectron attraction parameter $V=2$ (dotted lines) and $V=4$ (solid lines).}
\end{center}
\end{figure}
\begin{table}[h]
\label[{Table 1. Characteristic values of the chemical potential $\mu$ corresponding 
to the concentrations of charge carriers $n$ at temperatures $T=0$ and $T=T_c$ for the 
interelectron attraction parameter $V=2$.} 
\begin{center}
\begin{tabular}{cccccc}
\hline
\rule{0mm}{5mm}
$\mu$ & $-2.4$ & $-1.4$ & $0$ & $1.4$ & $2.4$\\
\hline
$n(T=0)\simeq$ & $0.396$ & $0.542$ & $0.798$ & $1.208$ & $1.703$\\ 
$n(T=T_c)\simeq$ & $0.397$ & $0.544$  & $0.800$ & $1.201$ & $1.742$\\ 
\hline
\end{tabular}
\end{center}
\end{table}
\begin{table}[h]
\label[{Table 2. Characteristic values of the chemical potential $\mu$ corresponding 
to the concentrations of charge carriers $n$ at temperatures $T=0$ and $T=T_c$ for the 
interelectron attraction parameter $V=4$.} 
\begin{center}
\begin{tabular}{cccccc}
\hline
\rule{0mm}{5mm}
$\mu$ & $-2.4$ & $-1.4$ & $0$ & $1.4$ & $2.4$\\
\hline
$n(T=0)\simeq$ & $0.413$ & $0.561$ & $0.836$ & $1.229$ & $1.577$\\ 
$n(T=T_c)\simeq$ & $0.410$ & $0.567$ & $0.842$ & $1.251$ & $1.677$\\ 
\hline
\end{tabular}
\end{center}
\end{table}

\section{Conclusion}

Within the framework of the self-consistent functional-integral theory, the influence of thermal 
fluctuations on the temperature behavior of the topological index $C_1$ of the chiral $d+id$ superconducting 
phase of a two-dimensional single-band model with effective attraction between electrons located at the 
sites of a triangular lattice within the second coordination sphere. We establish that when the  nodal 
points are located far inside (outside) the Fermi contour, taking into account thermal fluctuations leads 
to an expansion of the relative temperature ranges in which the values of the topological index are close 
to integer values $C_1=4 (-2)$. Furthermore, for the topological phases with $C_1=4$, the expansion 
of the relative temperature range is more effective. When the nodal points approach the Fermi contour, 
two distinct temperature dependence regimes are possible, depending on the magnitude of the effective 
attraction between the electrons. In the first case, at $V=2$, the regime of weak renormalization of the 
Fermi contour by thermal fluctuations is realized, and the temperature dependences of the TI, taking into 
account thermal fluctuations, differ slightly from those calculated in the HF approximation, and the 
topological crossover to values of $C_1=1$ is preserved. In the second case, in the regime of strong 
renormalization of the Fermi contour by thermal fluctuations, at $V=4$, there is no topological crossover 
to the values of $C_1=1$, but a new topological crossover occurs to the values of $C_1=-2$. Taking into 
account the thermal broadening  of the Fermi distribution leads to a pattern similar to the regime of weak 
renormalization of the Fermi contour. Assuming the bulk-boundary correspondence principle holds the 
detected effects in the system under consideration  should manifest in the behavior of edge states in a 
similar system with open boundaries. The results obtained could be used in technical devices of quantum 
thermal electronics.

\acknowledgments

This work was supported by the Ministry of Science and Higher Education 
of the Russian Federation (theme No. 124021900019-5).

\end{document}